\documentclass[twocolumn,prl,superscriptaddress,floatfix,showpacs]{revtex4-1}
\usepackage[utf8]{inputenc}
\usepackage{amsmath}
\usepackage{amssymb}
\usepackage{graphicx}

\makeatletter
\usepackage{graphics}
\usepackage{epsfig}
\usepackage{color}

\newcommand{\be}{\begin{equation}} \newcommand{\ee}{\end{equation}}
\newcommand{\bea}{\begin{eqnarray}} \newcommand{\eea}{\end{eqnarray}}
\DeclareMathOperator\arctanh{arctanh}

\newcommand{\beginsupplement}{
    \setcounter{section}{0}
    \renewcommand{\thesection}{S\arabic{section}}
    \setcounter{equation}{0}
    \renewcommand{\theequation}{S\arabic{equation}}
    \setcounter{figure}{0}
    \renewcommand{\thefigure}{S\arabic{figure}}}

\makeatother

\begin{document}

\title{Chaotic synchronization in adaptive networks of pulse-coupled oscillators}

\author{Germ\'an Mato} \affiliation{Medical Physics Department, Centro At\'omico Bariloche and Instituto Balseiro, 8400 San Carlos de Bariloche, R\'io Negro, Argentina} 
\affiliation{Max Planck Institut f\"ur Physik komplexer Systeme, N\"othnitzer Str. 38, 01187 Dresden, Germany}
\author{Antonio Politi}
\affiliation{Institute for Complex Systems and Mathematical Biology and Department of Physics, Aberdeen AB24 3UE, UK}
\affiliation{Max Planck Institut f\"ur Physik komplexer Systeme, N\"othnitzer Str. 38, 01187 Dresden, Germany}
\affiliation{CNR - Consiglio Nazionale delle Ricerche - Istituto dei Sistemi Complessi, via Madonna del Piano 10, 50019 Sesto Fiorentino, Italy}
\author{Alessandro Torcini} 
\affiliation{Laboratoire de Physique Th\'eorique et Mod\'elisation, CY Cergy Paris Universit\'e, CNRS, UMR 8089,
95302 Cergy-Pontoise cedex, France}
\affiliation{Max Planck Institut f\"ur Physik komplexer Systeme, N\"othnitzer Str. 38, 01187 Dresden, Germany}
\affiliation{CNR - Consiglio Nazionale delle Ricerche - Istituto dei Sistemi Complessi, via Madonna del Piano 10, 50019 Sesto Fiorentino, Italy}

\date{\today}

\begin{abstract}
Ensembles of phase-oscillators are known to exhibit a variety of collective regimes. 
Here, we show that a simple mean-field model involving two heterogenous populations of pulse-coupled oscillators, 
exhibits, in the strong-coupling limit, a robust irregular macroscopic dynamics.
The resulting, strongly synchronized, regime is sustained by a homeostatic mechanism 
induced by the shape of the phase-response curve combined with adaptive
coupling strength, included to account for energy dissipated by the pulse emission.
The proposed setup mimicks a neural network composed of excitatory and inhibitory
neurons.

\end{abstract}

\maketitle

Understanding the behavior of large ensembles of coupled oscillators is the objective of active research
since many years \cite{winfree,kuramoto2012}. 
This includes diverse areas such as engineering, 
neuroscience, and systems biology \cite{mirollo1990,liu1997,haken2006,dorfler2014}. One of the key points is the spontaneous emergence of various forms of synchronization \cite{pikovsky2001,strogatz2012}, i.e. of collective regimes out of given microscopic rules.
Identifying the relationship between the microscopic and macroscopic world is, however, non trivial and
a general theory is still lacking. In this Letter, we focus on the most challenging 
phenomenon: collective chaos (CC), a stochastic-like, partially shynchronized regime.
The emergence of CC in ensembles of oscillators which behave chaotically is a well established fact both
in heterogeneous~\cite{matthews1990} and homogenous \cite{kaneko1990,hakim1992,nakagawa1993} networks.

Pulse-coupled phase oscillators can give rise to quite complex scenarios.
Low dimensional CC has been reported in two coupled populations of spiking neurons ~\cite{olmi2011}, 
and of Winfree oscillators~\cite{pazo2014}, where it has been interpreted as a sort of chimera mechanism
\cite{abrams2004}. With reference to a single population, low-dimensional CC was found in quadratic 
integrate-and-fire (QIF) neurons in~\cite{pazo2016}, where the delay plays a crucial role, 
by increasing the phase space dimensionality.
Finally, high- (possibly infinite-) dimensional CC was reported in 
~\cite{luccioli2010,ullner2016}, where heterogeneity appears to be a crucial ingredient.

In all such cases, the connectivity is proportional (if not equal, as in mean-field models) to
the number of oscillators, while the coupling strength is assumed to scale as $1/N$, to avoid
unpleasant divergencies in the thermodynamic limit. We call this setup as S1.
However, there exists a second setup (S2), inherited form the study of spin glasses, 
where the coupling strength is assumed to be of order $1/\sqrt{N}$ \cite{sherrington1975,daido1992}: it makes sense whenever a number
of positive and negative coupling terms proportional to $N$ tend to cancel (balance) each other, suppressing divergencies.
The setup S2 is quite popular within neuroscience since the balance between excitatory and inhibitory
synaptic inputs spontaneously emerges in the presence of strong external currents \cite{bal1}, and very recently, it was proven
to work also when the currents are weak~\cite{politi2024}, if synaptic depression is included.
Besides, while the microscopic dynamics is almost regular in S1, it is strongly erratic
in S2, as expected for the brain neural activity~\cite{softky1992}.


In this Letter, we revisit the S1 setup, by studying the dynamics of two coupled populations of heterogeneous 
pulse-coupled phase oscillators (conveniently termed excitatory and inhibitory neurons). 
An important novelty is the inclusion of an adaptive mechanism -- short term depression (STD) -- proposed
in \cite{tsodyks1998,tsodyks2013}, which limits the firing activity of the neurons taking in account
the limitation of the resources. This mechanism gives rise 
to additional nonlinearities that persist in the infinite size limit without loosing microscopic variability, 
as shown by the mean field theory developed in \cite{mongillo2012}.

In our model, oscillator interactions are mediated by phase response curves (PRCs), as widely assumed in computational 
neuroscience~\cite{abbott1993,timme2002,denker2004,haken2006}.
Finally, we consider a globally coupled network, finding that above a critical value $G_\theta$ of the coupling
strength $G$, a fluctuationless asynchronous regime (AR) destabilizes, giving rise to CC accompanied by a highly
fluctuating microscopic dynamics.
The regime is kept in a balanced state by a self-adjusting synchronization which progressively enhances
upon increasing the coupling strength. This is the result of a non trivial synthesis of STD and
the PRC shape: an increased level of synchrony among the oscillators leads to a decrease in the effective interaction, 
because of the vanishing amplitude of the PRC in proximity to the pulse emission.

\paragraph*{Network Model.}

As shown in \cite{mirollo1990,van1996,timme2002,politi2024}, it is possible to derive analytically phase 
oscillator models describing the evolution of the membrane potentials 
of pulse-coupled supra-threshold neurons, whenever the neuron model is 
one dimensional.  In particular, by following \cite{politi2024} the evolution of the phase $\phi^{e/i}_n \in [0;1]$ of the $n$-th neuron
($e/i$ denotes the excitatory/inhibitory nature) can be written as 
\begin{equation}
	\dot \phi^{e/i}_n = \omega^{e/i}_n + G C^{e/i}Z(\phi^{e/i}_n) \qquad  1\le n \le N
\label{model0} 
\end{equation}
where the two populations are assumed to be characterized by the same number of neurons. 
Whenever the phase variable  $\phi^{e/i}_n$ reaches 1, an instantaneous pulse (a $\delta$-spike)
is delivered to all the excitatory and inhibitory oscillators and 
the phase is reset to zero.
The
natural frequencies $\omega^{e/i}_n$ are randomly selected, according to 
a generic distribution with a finite support $[\omega^{e/i}_m:\omega^{e/i}_M]$:
\begin{equation}
Q^{e/i}(\omega) = B^{e/i} \exp \left [ -(\omega-\omega^{e/i}_m)^{-1}(\omega^{e/i}_M-\omega)^{-1}\right ]
\label{freq} 
\end{equation}
where $B^{e/i}$ is a normalization factor \cite{note2}.
$G$ denotes the overall coupling constant: this is the main parameter we are going to tune
to explore the emergence of different regimes.

The function $Z(\phi)$ represents the PRC of each oscillator to a single external pulse. 
We consider the following PRC for both families of neurons 
\begin{equation}
  Z(\phi)= 16 \phi^2 (1-\phi)^2 \; ;
\end{equation}
this kind of PRC can be classified as type I \cite{hansel1995}, and it can be
put in correspondence with the dynamical behaviour of 
an excitable membrane of type I according to the Hodgkin classification \cite{canavier2006}.

Finally, $C^{e/i}$ denotes the aggregate input current stimulating excitatory/inhibitory neurons.
Under the assumption of a mean-field coupling, $C^{e/i}$ does not depend on the neuron label,
\begin{eqnarray}
C^{e/i} \equiv g^{e/i}_e E^{e/i} - g^{e/i}_i I \quad ,
\label{eq:C}
\end{eqnarray}
where the $g^{e/i}_{e/i}$ coefficients quantify the specific intra and inter (synaptic) coupling strengths 
of excitatory and inhibitory populations (here, we have set 
$g_e^e = 1$, $g_e^{i}= 1$, $g_{i}^e=1/2$, and $g_{i}^{i}=2$).
Furthermore, in Eq. \eqref{eq:C}
$E^{e/i}$ ($I$) represents the incoming (effective) excitatory (inhibitory) fields,
due to the previously delivered excitatory (inhibitory) pulses, namely
\begin{eqnarray}
	E^{e/i} &=& \frac{1}{N} \sum_{k, m|t^e(k,m)<t} x^{e/i}_k(t)  \delta(t-t^e(k,m)) \\
	I   &=& \frac{1}{N} \sum_{k, m|t^i(k,m)<t} \delta(t-t^i(k,m)) 
\label{field}
\end{eqnarray}
where $t^{e/i}(k,m)$ denotes the delivery time of the $m$-th spike by the $k$-th excitatory/inhibitory neuron.
$x^{e/i}_k \in[0,1]$ represents the synaptic efficacy of the $k$-th excitatory neuron. 
If the receiving neuron is inhibitory, $x^i_k \equiv 1$,
while excitatory-to-excitatory connections are characterized by an adaptive efficacy $x^e_k(t)$.
In particular, we assume that its evolution is controlled by an STD mechanism,
often invoked in neural systems to simulate the regulation of  excitatory activity \cite{tsodyks2013}. 
According to~\cite{tsodyks1998}, $x^e_k$ follows the equation 
\begin{equation}
        \dot x^e_k = \frac{(1-x^e_k)}{\tau_d}  -  u x^e_k \!\!\!\! \sum_{m|t^e(k,m)<t}\!\!\!\! \delta(t-t^e(k,m)) \; ,
\label{depr}
\end{equation}
where  $t^e(k,m)$ identifies the emission time of the $m$-th spike by the  $k$-th neuron itself.
Whenever the neuron spikes, its synaptic efficacy $x^e_k$ is reduced by a factor $(1-u)$,
representing the fraction of resources consumed to produce a post-synaptic spike.
So long as the $k$-th excitatory neuron does not spike, the variable $x^e_k$
increases towards 1 over a time scale $\tau_d$.
Altogether, there are $3N$ variables: $2N$ phases characterising the internal state of
all neurons, plus $N$ variables to quantify the instantaneous value of the excitatory-excitatory synaptic efficacy.

\paragraph*{Numerical Results.}
In this section, we provide a characterization of the network dynamics for the later interpretation.
We start by illustrating the collective behavior.
In Fig.~\ref{fig:fields}(a), we plot the
time averages of the three fields $I$, $E^{e/i}$ vs. the coupling strength $G$
for different network sizes from $N=8000$ to $N=32000$.
The fields appear to be asymptotic (essentially independent of $N$) and stay finite for increasing $G$
(they even show a slow decrease): a behavior suggestive of a balanced regime.
At $G= G_\theta \approx 13.5$, a dynamical phase transition occurs in the form of a Hopf bifurcation,
which separates an AR characterized by constant fields (below $G_\theta$), from a regime characterized
by oscillating fields (above $G_\theta$): the field standard deviations 
$\sigma_{e/i}$ are reported in the inset of Fig.~\ref{fig:fields}(a).

\begin{figure}
\begin{centering}
\includegraphics[width=0.5\textwidth,clip=true]{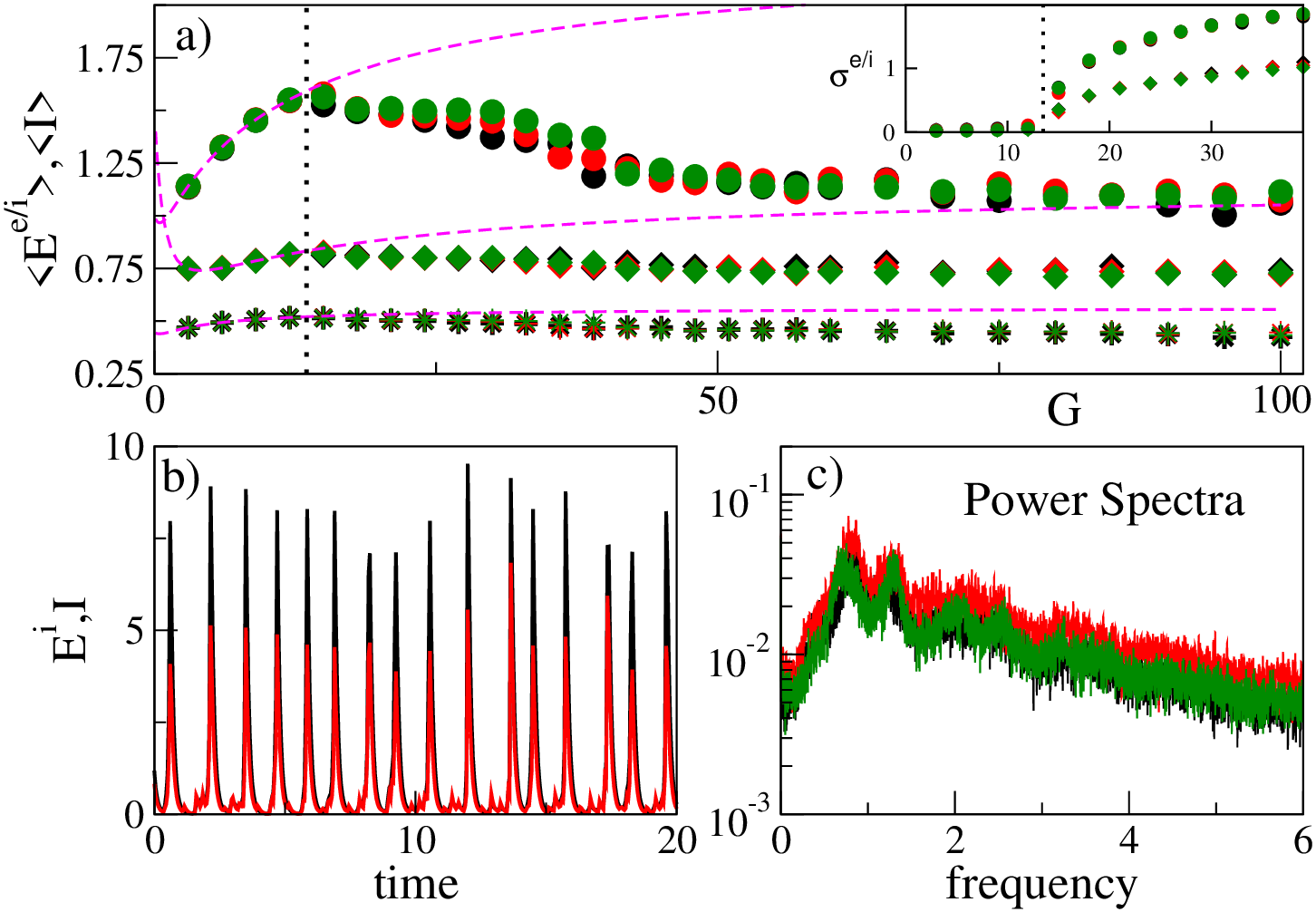}
\caption{(a) Time averaged synaptic excitatory $\langle E^{e/i} \rangle$ and inhibitory fields 
$\langle I \rangle$ versus $G$ for different system sizes. The circles denote $E^i$,
the diamond $I$ and the asterisks $E^e$, the dashed magenta lines refer to the exact results
for the AR and the dotted (black) vertical lines denote $G_\theta$. In the inset are reported the standard deviations $\sigma^{e/i}$ of the $E^i$ (circles) and $I$ fields (diamonds).
(b) Excitatory $E^i$ (black) and inhibitory $I$ (red) fields versus time. 
(c) Power spectra for the excitatory $E^i$ field for the quenched case
for different system sizes. Parameters $G=50$, the spectra refer to an integration time $t=5000$,
after discarding a transient of duration 500. 
The fields in (b-c) are filtered with
an exponential kernel for more details see \cite{supp}.
All the data refer to quenched cases. The considered sizes in (a) and (c) are denoted by different colors:
namely, $N=8000$ (black), $N=16000$ (red) and $N=32000$ (green). 
}
\label{fig:fields}
\end{centering}
\end{figure}

Direct evidence of the oscillatory regime can be appreciated in Fig.~\ref{fig:fields}(b) for $G=50$, 
where large irregular peaks are clearly visible in the evolution of the fields, suggesting the presence of a strong synchronization
of both excitatory and inhibitory neurons. 
A complementary view is finally presented in panel (c), where the Fourier power 
spectra of the excitatory field (those of the inhibitory fields
are similar) are plotted for different network sizes. 
There is no evident dependence on $N$, meaning that the
irregularity testified by the broadband structure of the spectra is also an asymptotic phenomenon.

We now shift the focus to the single-neuron level. Fig.~\ref{fig:behavior}(a) displays the firing rates of
the two populations for $G=50$: in both cases a plateau is visible, which reveals a mutual synchrony of
given groups of neurons at the self-determined frequency $\nu^{e/i} \approx 0.8$ ($N=16000$); a second plateau
at its second harmonic is also exhibited by the excitatory neurons.

So far, this phenomenology is very similar to that one reported in~\cite{luccioli2010,ullner2016}, 
where a heterogeneous network of exclusively inhibitory neurons was investigated.
However, the single-neuron behavior is, here, significantly more irregular.
The coefficients of variation ($CV$s) measuring the level of irregularity in the spiking activity of each
neuron~\cite{CV} instead of being at most around 0.1 as in~\cite{luccioli2010,ullner2016}, here they are
much larger and closer to the values measured in the brain cortex \cite{softky1993}, where the CV is around 1
(see Fig.~\ref{fig:behavior}(b)). This suggests and confirms that the simultaneous presence of excitation and inhibition is a 
crucial ingredient to obtain an irregular microscopic dynamics.

Once more, we wish to stress that 
the mean-field nature of the model implies that each (excitatory or inhibitory) neuron sees
the same field, so that the fluctuations are genuine collective properties rather than statistical
fluctuations  due to the ``sampling'' process in a sparse random network.

In order to test the robustness of this scenario, we have simulated also an {\it annealed} variant
of the {\it quenched} model, where the bare frequencies instead of being fixed forever, are 
randomly reset (according to the same distribution) every $N$ spikes. 
As shown in Fig. S1 in \cite{supp}, there are only minor quantitative variations such
as the bifurcation point which now occurs for $G_\theta \approx 10.5$.
At the microscopic level, in the annealed model, the neurons are necessarily characterized by the same firing rate
and the same $CV$: $\nu^e = 1.44 \pm 0.02$ and $\langle CV^e \rangle \simeq 0.43$ for the excitatory neurons;
$\nu^i = 0.85 \pm 0.03$ and $\langle CV^i \rangle \simeq 0.84$ for the inhibitory ones.
Interestingly, the $CV$'s are again large, showing that the same scenario emerges in a case (annealed setup) potentially
more amenable to an analytic treatment.

\begin{figure}
\begin{centering}
\includegraphics[width=0.5\textwidth,clip=true]{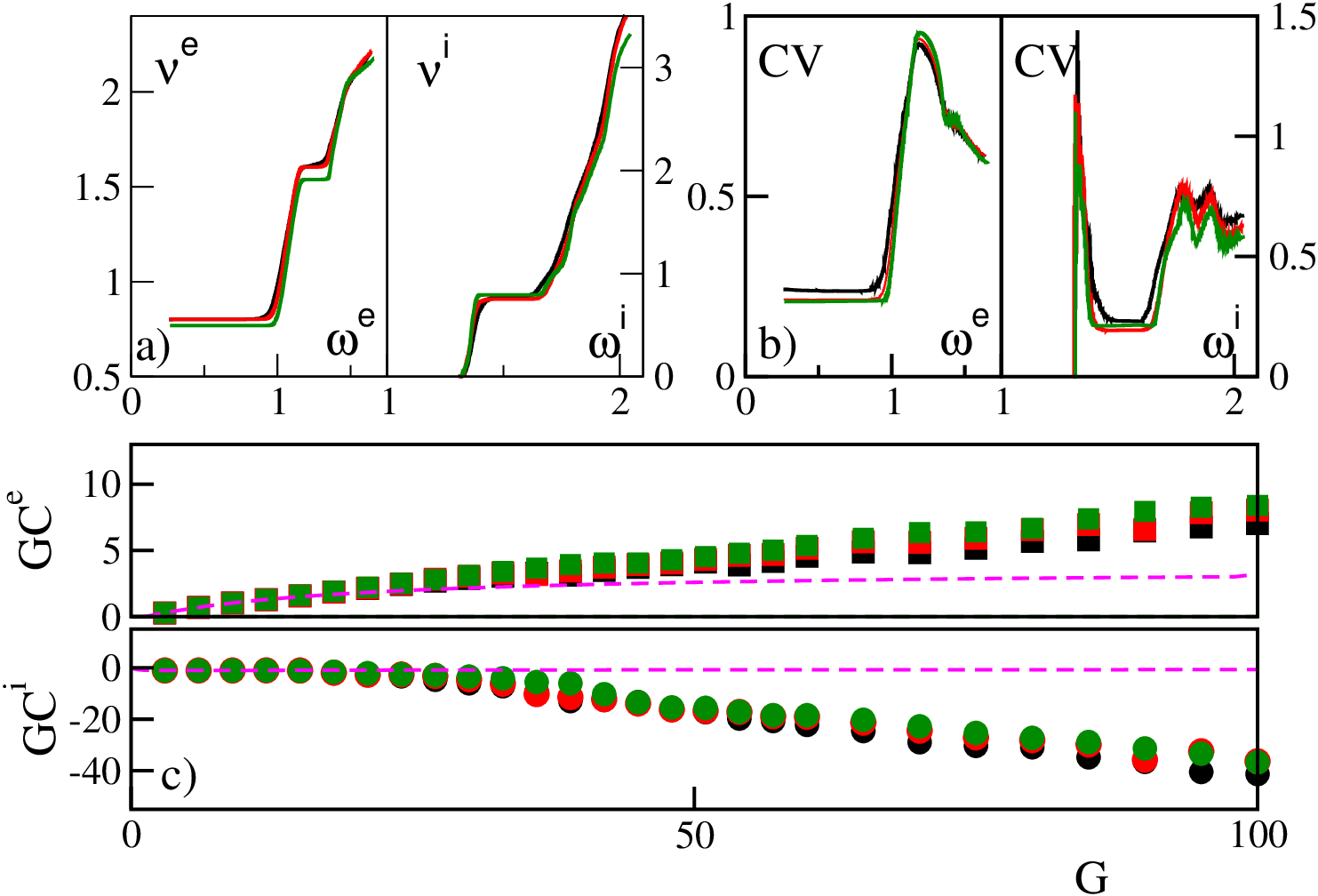}
\end{centering}
\caption{\label{fig:behavior} 
Average firing rates (a) and CV (b) versus their natural frequencies $\omega^{e/i}$
for excitatory (left panels) and inhibitory (right panels) neurons for different system sizes. 
The data refer to $G=50$. (c) Aggregate synaptic currents $C^{e}$ ($C^i$) multiplied by the coupling versus 
the synaptic coupling itself $G$ for excitatory (inhibitory) neurons for different system sizes are 
shown in the upper (lower) panel. The magenta dashed curves correspond to the exact calculations for the AR. All the data refer to quenched cases. The considered sizes are denoted by the same colors as in Fig. \ref{fig:fields} (a).
}
\end{figure}

\paragraph*{The Homeostatic Mechanism.}
We now address the question of how such a regime self-sustains, especially for large $G$ values.
We start by comparing the numerical results with the strictly constant AR, 
which can be determined from an exact mean field analysis.
The details of the calculations in the thermodynamic limit ($N\to\infty$)
are illustrated in the supplemental material~\cite{supp}; here we simply sketch
the procedure and present the final results.
Given two tentative values of the (constant) aggregate currents $C^{e/i}$, one can
determine the (constant) single-neuron firing rates by integrating the equations for
the phase and the synptic efficacy (see Eqs.~(\ref{model0},\ref{depr})).
Next, one can derive new $C^{e/i}$-estimates from the firing rates, using Eq.~(\ref{eq:C}):
they must be consistent with the initial assumptions.
In practice, the analysis amounts to searching a fixed point in the plane $(C^e,C^i)$,
and, once found, to determine the corresponding fields $E^{e/i}$, $I$ from the aggregate currents.
The self-consistent results for the fields $E^{e/i}$, $I$ are presented in Figs.~\ref{fig:fields}(a), 
(see the magenta dashed lines), where we see that they reproduce the numerical observations
below the threshold $G_\theta$, in agreement with the expectations, since we knew that this regime is asynchronous.
In the large $G$ limit, the AR still exists and the associated fields stay finite,
indicating that this regime is balanced. 
However, above $G_\theta$, the AR is no longer stable. The CC revealed by numerical simulations 
is also balanced, as visible in Fig.~\ref{fig:fields} (a), but
it is conceptually very different, not only because of the strong fluctuations of fields.
This can be appreciated in Fig.~\ref{fig:behavior}(c), where the scaled 
aggregate currents $G C^{e/i}$ are plotted versus the coupling $G$ itself:
they remain bounded in the AR (magenta dashed lines), while they diverge in the presence of CC (solid symbols). 
The asymptotic behavior of the AR can be explained as follows: the finitenes of $G C^{e/i}$ is a consequence
of the finiteness of the fields, combined with an asymmetry between the effective fields of order $1/G$.
This latter balance condition~\cite{scaling} can be satisfied at all 
because of the presence of synaptic depression, which makes $E^e$ 
smaller than $E^i$. In the absence of STD, the balanced AR could not exist, except for the special case $g^e_ig^i_e=g^e_eg^i_i$.

In the CC regime, the divergence of $GC^{e/i}$ visible in Fig.~\ref{fig:behavior}(c), seemingly contradicts the
finiteness of the fields seen in Fig.~\ref{fig:fields} (a). Consistency is restored by focusing on the role of the PRC. 
Typically, the efficacy of the incoming spikes, quantified by the PRC, is of order 1 irrespective of the coupling strength.
Here, the story is different as shown in Fig.~\ref{fig:3}(a), where we 
plot the time and ensemble average $\langle Z^i_c \rangle$ of the PRC, conditioned to the times of 
the spike arrivals (these are the only instants, where the PRC plays an active role in the neural dynamics).

There, we see that $\langle Z^i_c \rangle$ decreases as $1/G$ for large $G$ (see black circles)
thereby compensating the effect of the increasing coupling strength (in the presence of finite currents).
This is an indirect, though clear, indication of an underlying synchronization, since $Z(\phi)$ is very small only in the vicinity of the threshold and the reset.
Given the phase-like nature of the neuronal variable, one can quantify the degree of synchronization in terms of the modulus of the complex Kuramoto order parameter \cite{kuramoto2012} for the two populations,
defined as $ R^{e/i} \equiv \frac{1}{N} \left| \sum_{n=1}^N \mathrm{e}^{j\phi^{e/i}_n} \right|$, 
where $j$ denotes the imaginary unit.
In Fig.~\ref{fig:3}(a) we show the time averages $\langle R^{e/i}_c \rangle$ 
of the order parameters, conditioned to spike emissions, for the two
neural populations. In both cases, the parameter approaches 1 with a rate $1/G$, suggesting a convergence 
towards perfect synchronization when $G\to +\infty$.
However, this is not the whole story. In fact, the two unconditioned time averages 
$\langle R^{e/i} \rangle$  plotted in 
Fig.~\ref{fig:3}(b) remain strictly smaller than 1. 
These two observations thus reveal that the $G\to \infty$ limit is very peculiar. 
The probability distribution densities (PDFs) of the $R^i$ values, plotted in Fig.~\ref{fig:3}(c),
reveal that a power-law singularity emerges at $R^i=1$ for sufficiently large $G$.
Since the exponent approximately equal to 0.76, is smaller than
1, the singularity is integrable and does not even contribute to the average
value of $R^i$, which remains consistently smaller than 1.

\begin{figure}
\begin{centering}
\includegraphics[width=0.45\textwidth,clip=true]{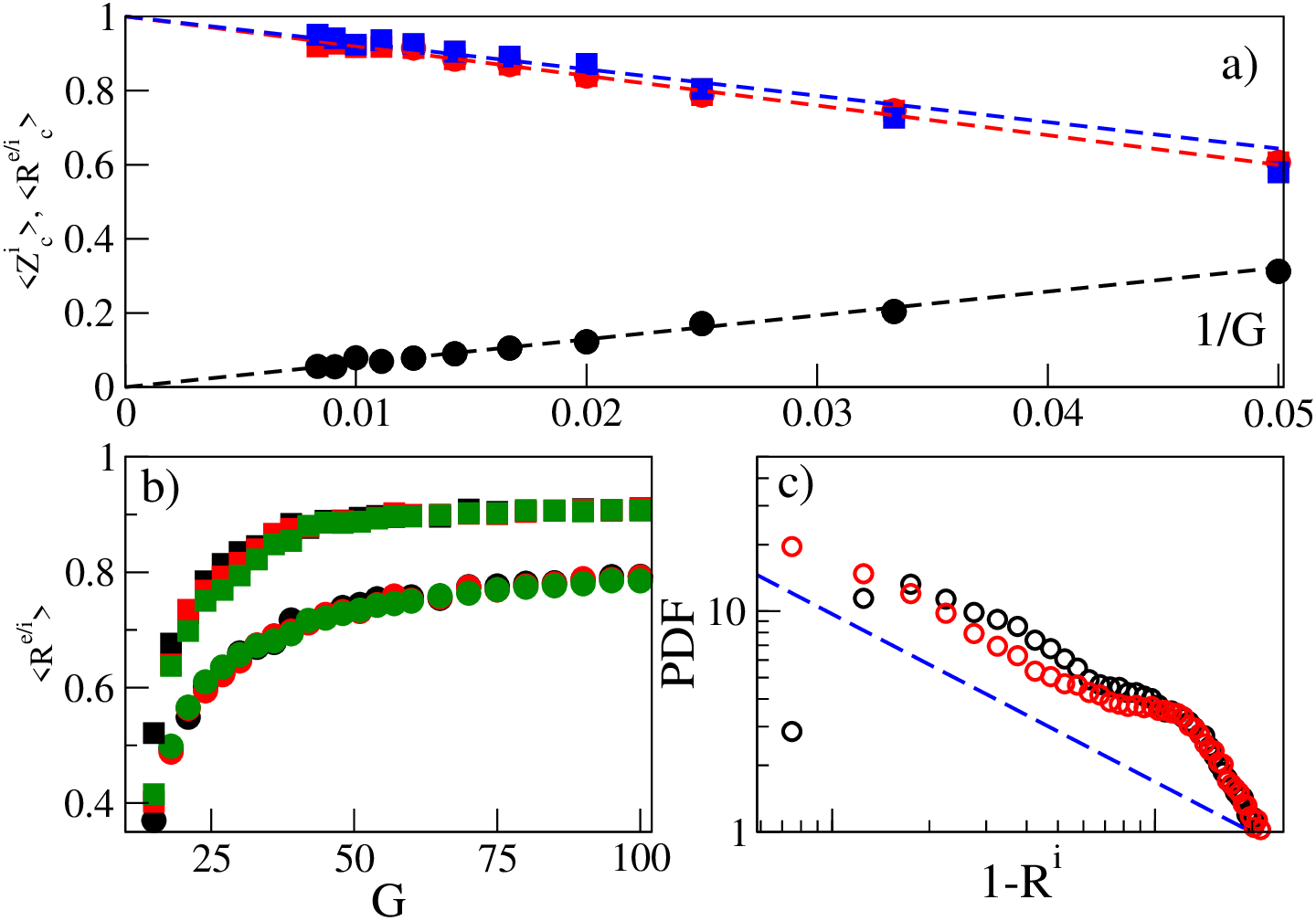}
\end{centering}
\caption{\label{fig:3} 
(a) Conditional PRC and order parameters: 
$\langle Z^i_c \rangle$ (black circles) $\langle R^e_c \rangle$ (red squares) and
$\langle R^i_c \rangle$ (blue squares) versus $1/G$.
We measure the average PRC of a typical inhibitory neuron whenever
an excitatory pulse reaches it and at the same time we measure $R^e$ and $R^i$.
The data refer to  $N=16000$ and to an average over 250000 events. 
(b)  Average value of the Kuramoto order parameter for the excitatory (circles) and inhibitory (squares)
population versus $G$ for various systems sizes: the sizes are encoded as in Fig. \ref{fig:fields} (a).
(c) PDF of $1-R^i$ for $G=50$ (black circles) and $G=100$ (red circles), the dashed
blue line denotes a power-law decay with exponent $0.76$. The data correspond to $N=16000$.
Parameters $K=0.5$.}
\end{figure}

Finally, we return to the fluctuations of the spiking activity. Whenever the underlying regime is
asynchronous, a large CV means that the neurons operate on average below threshold.
The membrane potential (or, equivalently, the phase) is confined to a valley and cannot reach 
the threshold: this can happen only when a strong fluctuation allows overcoming or removing 
the barrier. In the present case, the scenario is different: from Fig.~\ref{fig:behavior}(c), we see that $GC^e$ ($GC^i$)
becomes increasingly large and positive (negative) when $G$ is increased, suggesting that an
excitatory (inhibitory) neuron should find itself much above (below) threshold.
However this is not the case, since the large synaptic current values $G |C^{e/i}|$
typically contribute to the neuron dynamics at the spike arrivals, when the PRC is small, 
so that these terms do not sensibly affect the neuron activity.

\paragraph*{Conclusions.}
We have shown that a mean-field model of pulse-coupled (excitatory and inhibitory) oscillators may give rise
a strongly irregular microscopic and macroscopic regime.
This is at variance with typical balanced states, where the sparseness contributes to a stochastic-like
dynamics by enhancing statistical fluctuations; it is also different from the collective chaos exhibited
by chaotic units, as phase-oscillators are never chaotic, even under an external forcing.
We claim that the underlying mechanism originates from the selection of a suitable (type-I) PRC shape, which
induces a strong but imperfect synchronization. On the one hand, the synchronization stabilizes a balanced regime
via a homeostatic mechanism which desentisizes neurons when subject to strong spike bursts. On the other hand,
the PRC shape contributes to amplity the tiny differences due to the heterogeneous distribution of bare frequencies,
somehow mimicking the sensitivity of standard chaos.
The mean-field nature of the model, plus the evidence that the same scenario is observed in an annealed version
of the setup (as shown in \cite{supp}), together suggest the possibility to develop a theory able to justify our numerical observations.
This is the main goal of future work.

\begin{acknowledgments}
 
AT has received partial support by CY Generations (Grant No ANR-21-EXES-0008)
and by the Labex MME-DII (Grant No ANR-11-LBX-0023-01) all part of the French programme “Investissements
d’Avenir”. The work has been mainly realized at the Max Planck Institute for the Physics of Complex Systems (Dresden, Germany) as part of the activity of the Advanced Study Group 2016/17 ``From Microscopic to Collective Dynamics in Neural Circuits”. 
\end{acknowledgments}

  

\begin{thebibliography}{37}
\expandafter\ifx\csname natexlab\endcsname\relax\def\natexlab#1{#1}\fi
\expandafter\ifx\csname bibnamefont\endcsname\relax
  \def\bibnamefont#1{#1}\fi
\expandafter\ifx\csname bibfnamefont\endcsname\relax
  \def\bibfnamefont#1{#1}\fi
\expandafter\ifx\csname citenamefont\endcsname\relax
  \def\citenamefont#1{#1}\fi
\expandafter\ifx\csname url\endcsname\relax
  \def\url#1{\texttt{#1}}\fi
\expandafter\ifx\csname urlprefix\endcsname\relax\def\urlprefix{URL }\fi
\providecommand{\bibinfo}[2]{#2}
\providecommand{\eprint}[2][]{\url{#2}}

\bibitem[{\citenamefont{Winfree}(2001)}]{winfree}
\bibinfo{author}{\bibfnamefont{A.~T.} \bibnamefont{Winfree}},
  \emph{\bibinfo{title}{The Geometry of Biological Time}},
  vol.~\bibinfo{volume}{12} of \emph{\bibinfo{series}{Interdisciplinary Applied
  Mathematics}} (\bibinfo{publisher}{Springer-Verlag New York},
  \bibinfo{year}{2001}), \bibinfo{edition}{2nd} ed.

\bibitem[{\citenamefont{Kuramoto}(2012)}]{kuramoto2012}
\bibinfo{author}{\bibfnamefont{Y.}~\bibnamefont{Kuramoto}},
  \emph{\bibinfo{title}{Chemical oscillations, waves, and turbulence}},
  vol.~\bibinfo{volume}{19} (\bibinfo{publisher}{Springer Science \& Business
  Media}, \bibinfo{year}{2012}).

\bibitem[{\citenamefont{Mirollo and Strogatz}(1990)}]{mirollo1990}
\bibinfo{author}{\bibfnamefont{R.~E.} \bibnamefont{Mirollo}} \bibnamefont{and}
  \bibinfo{author}{\bibfnamefont{S.~H.} \bibnamefont{Strogatz}},
  \bibinfo{journal}{SIAM Journal on Applied Mathematics}
  \textbf{\bibinfo{volume}{50}}, \bibinfo{pages}{1645} (\bibinfo{year}{1990}).

\bibitem[{\citenamefont{Liu et~al.}(1997)\citenamefont{Liu, Weaver, Strogatz,
  and Reppert}}]{liu1997}
\bibinfo{author}{\bibfnamefont{C.}~\bibnamefont{Liu}},
  \bibinfo{author}{\bibfnamefont{D.~R.} \bibnamefont{Weaver}},
  \bibinfo{author}{\bibfnamefont{S.~H.} \bibnamefont{Strogatz}},
  \bibnamefont{and} \bibinfo{author}{\bibfnamefont{S.~M.}
  \bibnamefont{Reppert}}, \bibinfo{journal}{Cell}
  \textbf{\bibinfo{volume}{91}}, \bibinfo{pages}{855} (\bibinfo{year}{1997}).

\bibitem[{\citenamefont{Haken}(2006)}]{haken2006}
\bibinfo{author}{\bibfnamefont{H.}~\bibnamefont{Haken}},
  \emph{\bibinfo{title}{Brain dynamics: synchronization and activity patterns
  in pulse-coupled neural nets with delays and noise}}
  (\bibinfo{publisher}{Springer Science \& Business Media},
  \bibinfo{year}{2006}).

\bibitem[{\citenamefont{D{\"o}rfler and Bullo}(2014)}]{dorfler2014}
\bibinfo{author}{\bibfnamefont{F.}~\bibnamefont{D{\"o}rfler}} \bibnamefont{and}
  \bibinfo{author}{\bibfnamefont{F.}~\bibnamefont{Bullo}},
  \bibinfo{journal}{Automatica} \textbf{\bibinfo{volume}{50}},
  \bibinfo{pages}{1539} (\bibinfo{year}{2014}).

\bibitem[{\citenamefont{Pikovsky et~al.}(2001)\citenamefont{Pikovsky,
  Rosenblum, Kurths, and Synchronization}}]{pikovsky2001}
\bibinfo{author}{\bibfnamefont{A.}~\bibnamefont{Pikovsky}},
  \bibinfo{author}{\bibfnamefont{M.}~\bibnamefont{Rosenblum}},
  \bibinfo{author}{\bibfnamefont{J.}~\bibnamefont{Kurths}}, \bibnamefont{and}
  \bibinfo{author}{\bibfnamefont{A.}~\bibnamefont{Synchronization}},
  \bibinfo{journal}{Self} \textbf{\bibinfo{volume}{2}}, \bibinfo{pages}{3}
  (\bibinfo{year}{2001}).

\bibitem[{\citenamefont{Strogatz}(2012)}]{strogatz2012}
\bibinfo{author}{\bibfnamefont{S.~H.} \bibnamefont{Strogatz}},
  \emph{\bibinfo{title}{Sync: How order emerges from chaos in the universe,
  nature, and daily life}} (\bibinfo{publisher}{Hachette UK},
  \bibinfo{year}{2012}).

\bibitem[{\citenamefont{Matthews and Strogatz}(1990)}]{matthews1990}
\bibinfo{author}{\bibfnamefont{P.~C.} \bibnamefont{Matthews}} \bibnamefont{and}
  \bibinfo{author}{\bibfnamefont{S.~H.} \bibnamefont{Strogatz}},
  \bibinfo{journal}{Physical review letters} \textbf{\bibinfo{volume}{65}},
  \bibinfo{pages}{1701} (\bibinfo{year}{1990}).

\bibitem[{\citenamefont{Kaneko}(1990)}]{kaneko1990}
\bibinfo{author}{\bibfnamefont{K.}~\bibnamefont{Kaneko}},
  \bibinfo{journal}{Physical review letters} \textbf{\bibinfo{volume}{65}},
  \bibinfo{pages}{1391} (\bibinfo{year}{1990}).

\bibitem[{\citenamefont{Hakim and Rappel}(1992)}]{hakim1992}
\bibinfo{author}{\bibfnamefont{V.}~\bibnamefont{Hakim}} \bibnamefont{and}
  \bibinfo{author}{\bibfnamefont{W.-J.} \bibnamefont{Rappel}},
  \bibinfo{journal}{Physical Review A} \textbf{\bibinfo{volume}{46}},
  \bibinfo{pages}{R7347} (\bibinfo{year}{1992}).

\bibitem[{\citenamefont{Nakagawa and Kuramoto}(1993)}]{nakagawa1993}
\bibinfo{author}{\bibfnamefont{N.}~\bibnamefont{Nakagawa}} \bibnamefont{and}
  \bibinfo{author}{\bibfnamefont{Y.}~\bibnamefont{Kuramoto}},
  \bibinfo{journal}{Progress of Theoretical Physics}
  \textbf{\bibinfo{volume}{89}}, \bibinfo{pages}{313} (\bibinfo{year}{1993}).

\bibitem[{\citenamefont{Olmi et~al.}(2011)\citenamefont{Olmi, Politi, and
  Torcini}}]{olmi2011}
\bibinfo{author}{\bibfnamefont{S.}~\bibnamefont{Olmi}},
  \bibinfo{author}{\bibfnamefont{A.}~\bibnamefont{Politi}}, \bibnamefont{and}
  \bibinfo{author}{\bibfnamefont{A.}~\bibnamefont{Torcini}},
  \bibinfo{journal}{Europhysics Letters} \textbf{\bibinfo{volume}{92}},
  \bibinfo{pages}{60007} (\bibinfo{year}{2011}).

\bibitem[{\citenamefont{Paz{\'o} and Montbri{\'o}}(2014)}]{pazo2014}
\bibinfo{author}{\bibfnamefont{D.}~\bibnamefont{Paz{\'o}}} \bibnamefont{and}
  \bibinfo{author}{\bibfnamefont{E.}~\bibnamefont{Montbri{\'o}}},
  \bibinfo{journal}{Physical Review X} \textbf{\bibinfo{volume}{4}},
  \bibinfo{pages}{011009} (\bibinfo{year}{2014}).

\bibitem[{\citenamefont{Abrams and Strogatz}(2004)}]{abrams2004}
\bibinfo{author}{\bibfnamefont{D.~M.} \bibnamefont{Abrams}} \bibnamefont{and}
  \bibinfo{author}{\bibfnamefont{S.~H.} \bibnamefont{Strogatz}},
  \bibinfo{journal}{Physical review letters} \textbf{\bibinfo{volume}{93}},
  \bibinfo{pages}{174102} (\bibinfo{year}{2004}).

\bibitem[{\citenamefont{Paz{\'o} and Montbri{\'o}}(2016)}]{pazo2016}
\bibinfo{author}{\bibfnamefont{D.}~\bibnamefont{Paz{\'o}}} \bibnamefont{and}
  \bibinfo{author}{\bibfnamefont{E.}~\bibnamefont{Montbri{\'o}}},
  \bibinfo{journal}{Physical review letters} \textbf{\bibinfo{volume}{116}},
  \bibinfo{pages}{238101} (\bibinfo{year}{2016}).

\bibitem[{\citenamefont{Luccioli and Politi}(2010)}]{luccioli2010}
\bibinfo{author}{\bibfnamefont{S.}~\bibnamefont{Luccioli}} \bibnamefont{and}
  \bibinfo{author}{\bibfnamefont{A.}~\bibnamefont{Politi}},
  \bibinfo{journal}{Phys. Rev. Lett.} \textbf{\bibinfo{volume}{105}},
  \bibinfo{pages}{158104} (\bibinfo{year}{2010}).

\bibitem[{\citenamefont{Ullner and Politi}(2016)}]{ullner2016}
\bibinfo{author}{\bibfnamefont{E.}~\bibnamefont{Ullner}} \bibnamefont{and}
  \bibinfo{author}{\bibfnamefont{A.}~\bibnamefont{Politi}},
  \bibinfo{journal}{Physical Review X} \textbf{\bibinfo{volume}{6}},
  \bibinfo{pages}{011015} (\bibinfo{year}{2016}).

\bibitem[{\citenamefont{Sherrington and Kirkpatrick}(1975)}]{sherrington1975}
\bibinfo{author}{\bibfnamefont{D.}~\bibnamefont{Sherrington}} \bibnamefont{and}
  \bibinfo{author}{\bibfnamefont{S.}~\bibnamefont{Kirkpatrick}},
  \bibinfo{journal}{Physical review letters} \textbf{\bibinfo{volume}{35}},
  \bibinfo{pages}{1792} (\bibinfo{year}{1975}).

\bibitem[{\citenamefont{Daido}(1992)}]{daido1992}
\bibinfo{author}{\bibfnamefont{H.}~\bibnamefont{Daido}},
  \bibinfo{journal}{Physical review letters} \textbf{\bibinfo{volume}{68}},
  \bibinfo{pages}{1073} (\bibinfo{year}{1992}).

\bibitem[{\citenamefont{van Vreeswijk and Sompolinsky}(1996)}]{bal1}
\bibinfo{author}{\bibfnamefont{C.}~\bibnamefont{van Vreeswijk}}
  \bibnamefont{and}
  \bibinfo{author}{\bibfnamefont{H.}~\bibnamefont{Sompolinsky}},
  \bibinfo{journal}{Science} \textbf{\bibinfo{volume}{274}},
  \bibinfo{pages}{1724} (\bibinfo{year}{1996}).

\bibitem[{\citenamefont{Politi and Torcini}(2024)}]{politi2024}
\bibinfo{author}{\bibfnamefont{A.}~\bibnamefont{Politi}} \bibnamefont{and}
  \bibinfo{author}{\bibfnamefont{A.}~\bibnamefont{Torcini}},
  \bibinfo{journal}{Chaos: An Interdisciplinary Journal of Nonlinear Science}
  \textbf{\bibinfo{volume}{34}} (\bibinfo{year}{2024}).

\bibitem[{\citenamefont{Softky and Koch}(1992)}]{softky1992}
\bibinfo{author}{\bibfnamefont{W.~R.} \bibnamefont{Softky}} \bibnamefont{and}
  \bibinfo{author}{\bibfnamefont{C.}~\bibnamefont{Koch}}
  (\bibinfo{year}{1992}).

\bibitem[{\citenamefont{Tsodyks et~al.}(1998)\citenamefont{Tsodyks, Pawelzik,
  and Markram}}]{tsodyks1998}
\bibinfo{author}{\bibfnamefont{M.}~\bibnamefont{Tsodyks}},
  \bibinfo{author}{\bibfnamefont{K.}~\bibnamefont{Pawelzik}}, \bibnamefont{and}
  \bibinfo{author}{\bibfnamefont{H.}~\bibnamefont{Markram}},
  \bibinfo{journal}{Neural computation} \textbf{\bibinfo{volume}{10}},
  \bibinfo{pages}{821} (\bibinfo{year}{1998}).

\bibitem[{\citenamefont{Tsodyks and Wu}(2013)}]{tsodyks2013}
\bibinfo{author}{\bibfnamefont{M.}~\bibnamefont{Tsodyks}} \bibnamefont{and}
  \bibinfo{author}{\bibfnamefont{S.}~\bibnamefont{Wu}},
  \bibinfo{journal}{Scholarpedia} \textbf{\bibinfo{volume}{8}},
  \bibinfo{pages}{3153} (\bibinfo{year}{2013}).

\bibitem[{\citenamefont{Mongillo et~al.}(2012)\citenamefont{Mongillo, Hansel,
  and Van~Vreeswijk}}]{mongillo2012}
\bibinfo{author}{\bibfnamefont{G.}~\bibnamefont{Mongillo}},
  \bibinfo{author}{\bibfnamefont{D.}~\bibnamefont{Hansel}}, \bibnamefont{and}
  \bibinfo{author}{\bibfnamefont{C.}~\bibnamefont{Van~Vreeswijk}},
  \bibinfo{journal}{Physical review letters} \textbf{\bibinfo{volume}{108}},
  \bibinfo{pages}{158101} (\bibinfo{year}{2012}).

\bibitem[{\citenamefont{Abbott and Van~Vreeswijk}(1993)}]{abbott1993}
\bibinfo{author}{\bibfnamefont{L.~F.} \bibnamefont{Abbott}} \bibnamefont{and}
  \bibinfo{author}{\bibfnamefont{C.}~\bibnamefont{Van~Vreeswijk}},
  \bibinfo{journal}{Physical Review E} \textbf{\bibinfo{volume}{48}},
  \bibinfo{pages}{1483} (\bibinfo{year}{1993}).

\bibitem[{\citenamefont{Timme et~al.}(2002)\citenamefont{Timme, Wolf, and
  Geisel}}]{timme2002}
\bibinfo{author}{\bibfnamefont{M.}~\bibnamefont{Timme}},
  \bibinfo{author}{\bibfnamefont{F.}~\bibnamefont{Wolf}}, \bibnamefont{and}
  \bibinfo{author}{\bibfnamefont{T.}~\bibnamefont{Geisel}},
  \bibinfo{journal}{Physical review letters} \textbf{\bibinfo{volume}{89}},
  \bibinfo{pages}{258701} (\bibinfo{year}{2002}).

\bibitem[{\citenamefont{Denker et~al.}(2004)\citenamefont{Denker, Timme,
  Diesmann, Wolf, and Geisel}}]{denker2004}
\bibinfo{author}{\bibfnamefont{M.}~\bibnamefont{Denker}},
  \bibinfo{author}{\bibfnamefont{M.}~\bibnamefont{Timme}},
  \bibinfo{author}{\bibfnamefont{M.}~\bibnamefont{Diesmann}},
  \bibinfo{author}{\bibfnamefont{F.}~\bibnamefont{Wolf}}, \bibnamefont{and}
  \bibinfo{author}{\bibfnamefont{T.}~\bibnamefont{Geisel}},
  \bibinfo{journal}{Physical review letters} \textbf{\bibinfo{volume}{92}},
  \bibinfo{pages}{074103} (\bibinfo{year}{2004}).

\bibitem[{\citenamefont{van Vreeswijk}(1996)}]{van1996}
\bibinfo{author}{\bibfnamefont{C.}~\bibnamefont{van Vreeswijk}},
  \bibinfo{journal}{Physical Review E} \textbf{\bibinfo{volume}{54}},
  \bibinfo{pages}{5522} (\bibinfo{year}{1996}).

\bibitem[{not()}]{note2}
\bibinfo{note}{Here, we have set $\omega^e_m = 0.1997$, $\omega^e_M=1.8003$
  ($B^e=5.5999\ldots$), and $\omega^i_m = 0.81$, $\omega^i_M=2.19$
  ($B^i=12.2359\ldots$).The two choices correspond to an average 1 and 1.5,
  respectively and have the same standard deviation as flat distributions of
  total width 1 and 0.8, respectively.}

\bibitem[{\citenamefont{Hansel et~al.}(1995)\citenamefont{Hansel, Mato, and
  Meunier}}]{hansel1995}
\bibinfo{author}{\bibfnamefont{D.}~\bibnamefont{Hansel}},
  \bibinfo{author}{\bibfnamefont{G.}~\bibnamefont{Mato}}, \bibnamefont{and}
  \bibinfo{author}{\bibfnamefont{C.}~\bibnamefont{Meunier}},
  \bibinfo{journal}{Neural computation} \textbf{\bibinfo{volume}{7}},
  \bibinfo{pages}{307} (\bibinfo{year}{1995}).

\bibitem[{\citenamefont{Canavier}(2006)}]{canavier2006}
\bibinfo{author}{\bibfnamefont{C.~C.} \bibnamefont{Canavier}},
  \bibinfo{journal}{Scholarpedia} \textbf{\bibinfo{volume}{1}},
  \bibinfo{pages}{1332} (\bibinfo{year}{2006}).

\bibitem[{sup()}]{supp}
\bibinfo{note}{See Supplemental Material at [URL will be inserted by publisher]
  for details on the self-consistent analysis for the asynchronous regime, for
  the results on the annealed case and details on the performed simulations.}

\bibitem[{CV()}]{CV}
\bibinfo{note}{The coefficient of variation $CV$ of a certain neuron is the
  ratio between the standard deviation and the mean of the inter-spike
  intervals associated to the train of spikes emitted by the considered neuron.
  $CV=0$ corresponds to a periodic activity, while $CV=1$ is observable for a
  Poissonian spike train.}

\bibitem[{\citenamefont{Softky and Koch}(1993)}]{softky1993}
\bibinfo{author}{\bibfnamefont{W.~R.} \bibnamefont{Softky}} \bibnamefont{and}
  \bibinfo{author}{\bibfnamefont{C.}~\bibnamefont{Koch}},
  \bibinfo{journal}{Journal of neuroscience} \textbf{\bibinfo{volume}{13}},
  \bibinfo{pages}{334} (\bibinfo{year}{1993}).

\bibitem[{sca()}]{scaling}
\bibinfo{note}{Usually in the dynamical balance theory \cite{bal1} the coupling
  strenght scales as $G \propto 1/\sqrt{N}$ therefore the difference of the
  effective fields, contained in $C^{e/i}$, grows as $G^{-1} \propto \sqrt{N}$
  to obtain the balancing.}

\end{thebibliography}

\title{Supplemental Material: \\ Chaotic synchronization in adaptive networks of pulse-coupled oscillators}

\beginsupplement

\newpage

\section*{Supplemental material}

\subsection{Asynchronous regime}
In the asynchronous regime and for $N \to\infty$, the neurons feel constant fields.
It is convenient to write their evolution equation as
\begin{equation}
 \dot \phi^{e/i} = \omega +  B^{e/i} Z(\phi) \; .
 \label{eqa:exin} 
\end{equation}
where $B^{e/i}= G C^{e/i}$ is defined as
\begin{equation}
	B^{e/i} = G\left [g^{e/i}_e  E^{e/i} -  g^{e/i}_i  I \right ]  \; .
 \label{eqa:AIE} 
\end{equation}
By integrating Eq.~(\ref{eqa:exin}), one can determine the interspike interval (ISI)
\begin{equation}
   T(\omega,B^{e/i}) \equiv  
\int_0^1  \frac{d \phi}{\omega + B^{e/i} Z(\phi)}  \; ,
 \label{eqa:in2}
\end{equation}
From the ISI, one can obtain the three fields.
$E^i$ and $I$ are obtained by averaging the inverse of the ISI over the corresponding distributions
\begin{equation}
E^i = \int d\omega \frac{Q^e (\omega) }{T(\omega,B^e)} \quad , \quad
I = \int d\omega \frac{Q^i (\omega) }{T(\omega,B^i)} \; .
\label{eqa:in3}
\end{equation}
The determination of $E^e$ requires the knowledge of the
the synaptic efficacy $x^e(\omega)$ at the spike-emission time,
\begin{equation}
E^e = \int d\omega \frac{Q^e(\omega) x^{e}(\omega,B^e)}{T(\omega,B^e)} \; .
\label{eqa:link3}
\end{equation}
The time-dependent efficacy $x^e(t)$ obeys the evolution equation
\begin{equation}
	\dot x^{e} = \gamma [1-x^{e}] \; .
\label{eqa:sd1}
\end{equation}
Its general solution is
\begin{equation}
x^{e}(t) = 1 - (1-x^{e}(0))\mathrm{e}^{-\gamma t } \; .
\label{eqa:sd1-sol}
\end{equation}
The initial condition $x^{e}(0)$ can be determined self-consistently, by imposing
$x^{e}(0) = (1-u) x^{e}(T^e)$, where $T^e = T(\omega,B^e)$, so that
\begin{equation}
x^{e}(\omega,B^e) = 1 - (1-(1-u) x^{e}(\omega,B^e))\mathrm{e}^{-\gamma T^e } \; .
\label{eqa:sd2a}
\end{equation}
whose explicit solution is
\begin{equation}
x^{e}(\omega,B^e) = \frac{1- \mathrm{e}^{-\gamma T^e }}{1-(1-u) \mathrm{e}^{-\gamma T^e }} 
\label{eqa:sd2}
\end{equation}
By inserting this expression into Eq.~(\ref{eqa:link3}), one can finally determine the last field $E^e$.

Altogether, given $B^i$ and $B^e$, one can determine $E^e$, $E^i$ and $I$.
Therefore, we can interpret Eq.~(\ref{eqa:AIE}) as two self-consistent conditions
for two unknowns. In other words, the problem of determining the asynchronous state amounts
to finding a fixed point in a two-dimensional space.

\subsubsection{Numerics}

If we assume that the PRC is $Z_2(x)= 16 x^2(1-x)^2$, the interspike interval 
Eq.~(\ref{eqa:in2}) can be written as (here we drop the superscript for the sake of simplicity)
\begin{equation}
    T(\omega,B) 
  = \int_0^1 \frac{d\phi}{\omega + 16 B \phi^2(1-\phi)^2} 
    = \frac{1}{|B|} \int_{0}^{1} \frac{dv}{a^4 \pm [1-v^2]^2} \; ,
 \label{eq:intin1} 
\end{equation}
where we have introduced the new variable $v= 2\phi -1$,
while $a^4= \omega/|B|$, and the sign in the denominator is the sign of $B$.

The integral can be solved analytically.
If the sign is negative, the result is
\begin{equation}
    T(\omega,B) = \frac{1}{2a^2|B|}
\left [
\frac{\arctan (a^2-1)^{-1/2}}{(a^2-1)^{1/2}} +
\frac{\arctanh (a^2+1)^{-1/2}}{(a^2+1)^{1/2}} 
\right ]
 \label{eq:intin2a} 
\end{equation}
Notice that $a<1$ means that the slowing down of the coupling is so strong as to block the neuron
from firing. Hence, the corresponding neuron does not contribute to the firing rate.

If the sign is positive, the result is
\begin{equation}
    T(\omega,B) = -\frac{j}{2a^2|B|}
\left [
u \arctan u + v \arctan v 
\right ]
 \label{eq:intin2b} 
\end{equation}
where
\begin{eqnarray}
u &=& \mathrm{e}^{i\pi/4}(a^2-j)^{-1/2}\\
v &=& \mathrm{e}^{3i\pi/4}(a^2+j)^{1/2} 
\label{eq:intin2c} 
\end{eqnarray}
and $j$ is the imaginary unit. 

\subsubsection{Large coupling limit}
In the large $G$ limit, a bounded solution can be obtained if the prefactor of $G$
in Eq.~(\ref{eqa:AIE}) vanishes. The resulting equations imply that, at leading order,
\begin{eqnarray}
        \frac{ E^{e}}{E^i} &=&  \frac{g^{e}_i g^i_e}{g^e_eg^i_i}  \label{eqa:fin1}  \\
        I &=&  \frac{g^{i}_e}{g^i_i} E^i   \label{eqa:fin2} 
\end{eqnarray}
As from Eqs.~(\ref{eqa:in3},\ref{eqa:link3}), both $E^e$ and $E^i$ are function only of $B^e$, 
which can thus be determined by imposing the condition (\ref{eqa:fin1}).
Once $B^e$ is estimated, one can determine $B^i$ from Eqs.~(\ref{eqa:fin2},\ref{eqa:in3}).

For the parameters employed in the Letter and $u = 0.50$ and $\gamma = 0.35$
we have found $ B^+ = 3.628574  $ and $B^- \simeq 0.619201$
and $E^- \simeq 2.256595$ and $I \simeq 1.128311$.

\subsection{Annealed Disorder}

In this Section we report the results for the annealed case. The natural frequencies $\{\omega^{e/i}\}$
are still selected from the distribution reported in Eq. (2) of the Letter, but their values,
are randomly updated every $N$ spikes of the whole network.

\begin{figure}
\begin{centering}
\includegraphics[width=0.5\textwidth,clip=true]{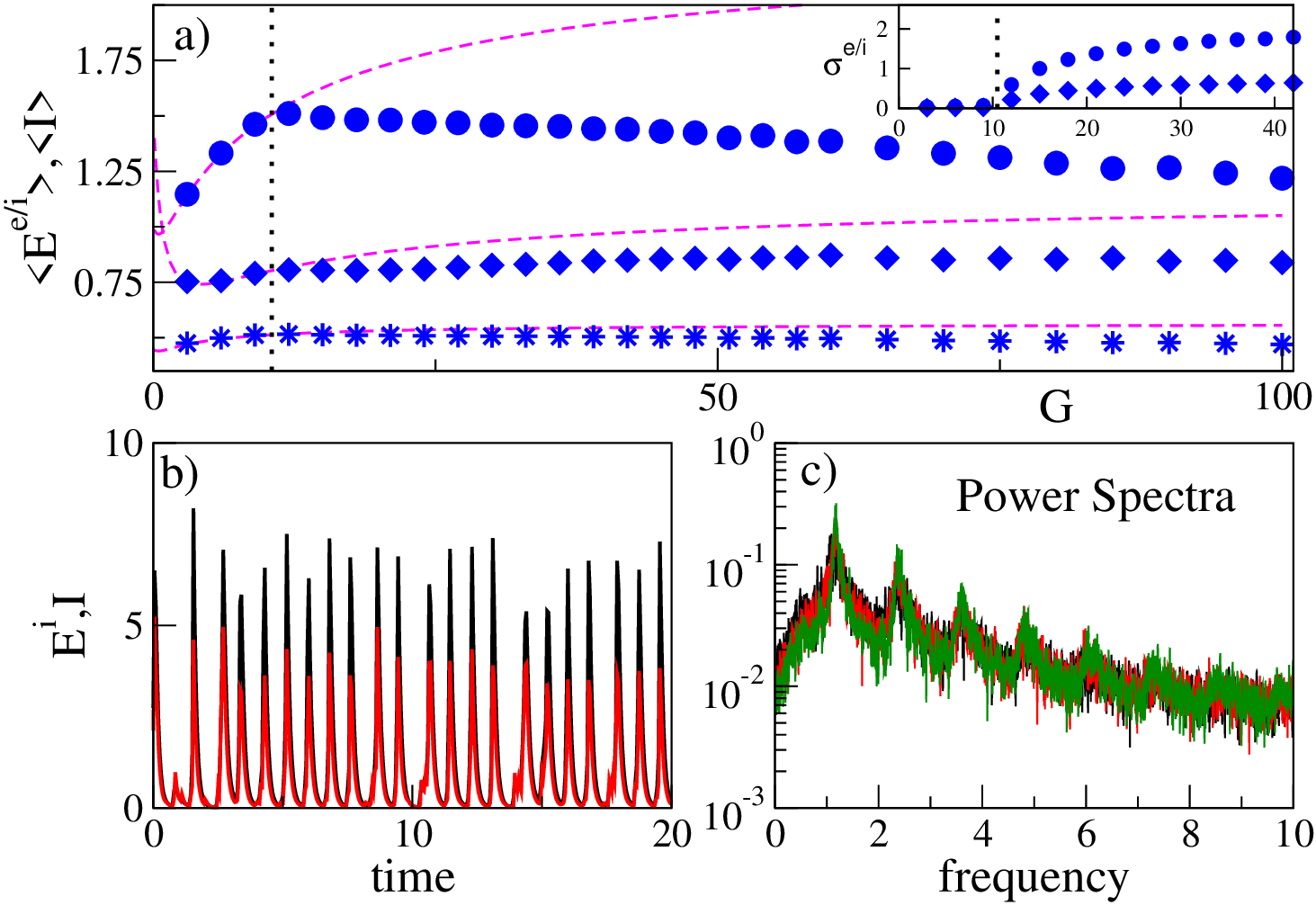}
\caption{(a) Time averaged synaptic excitatory $\langle E^{e/i} \rangle$ and inhibitory fields 
$\langle I \rangle$ versus $G$ for $N=16000$. The circles denote $E^i$,
the diamond $I$ and the asterisks $E^e$, the dashed (magenta) lines refer to the exact mean field
results for the asynchronous regime. In the inset are reported the standard deviation of the $E^i$ (circles) and $I$ fields
(diamonds). (b) Excitatory $E^i$ (black) and inhibitory $I$ (red) fields versus time. 
(c) Power spectra for the excitatory $E^i$ field for different system sizes:
$N=8000$ (black), $N=16000$ (red) and $N=32000$ (green). 
Parameters $G=50$ the spectra refer to an integration time $t=5000$, after discarding a transient 
of duration 500.
The fields in (b-c) are exponentially filtered with $\alpha=10$. All the data refer to the annealed case.  
}
\label{fig:fields}
\end{centering}
\end{figure}

In particular, in Fig.~\ref{fig:fields}(a), we plot the
time averages of the three fields $I$, $E^{e/i}$ versus the coupling strength $G$
for different network sizes from $N=8000$ to $N=32000$.
The field values appear to be essentially independent of $N$ and remain finite for increasing $G$:
a behavior consistent with a balanced regime.
At $G= G_\theta \approx 10.5$, we observe a Hopf bifurcation separating an AR characterized by constant fields (below $G_\theta$), from a regime characterized by oscillating fields (above $G_\theta$). As clearly visible from the field standard deviations 
$\sigma_{e/i}$ reported in the inset of Fig.~\ref{fig:fields}(a).

Direct evidence of the oscillatory regime can be appreciated in Fig.~\ref{fig:fields}(b) for $G=50$, 
where large irregular peaks are clearly visible in the evolution of the fields, suggesting the presence of a strong synchronization
of both excitatory and inhibitory neurons. 
A complementary view is finally presented in panel (c), where the Fourier power 
spectra of the excitatory field are reported for different network sizes. 
There is no evident dependence on $N$, meaning that the
irregularity testified by the broadband structure of the spectra is also an asymptotic phenomenon.

\begin{figure}
\begin{centering}
\includegraphics[width=0.5\textwidth,clip=true]{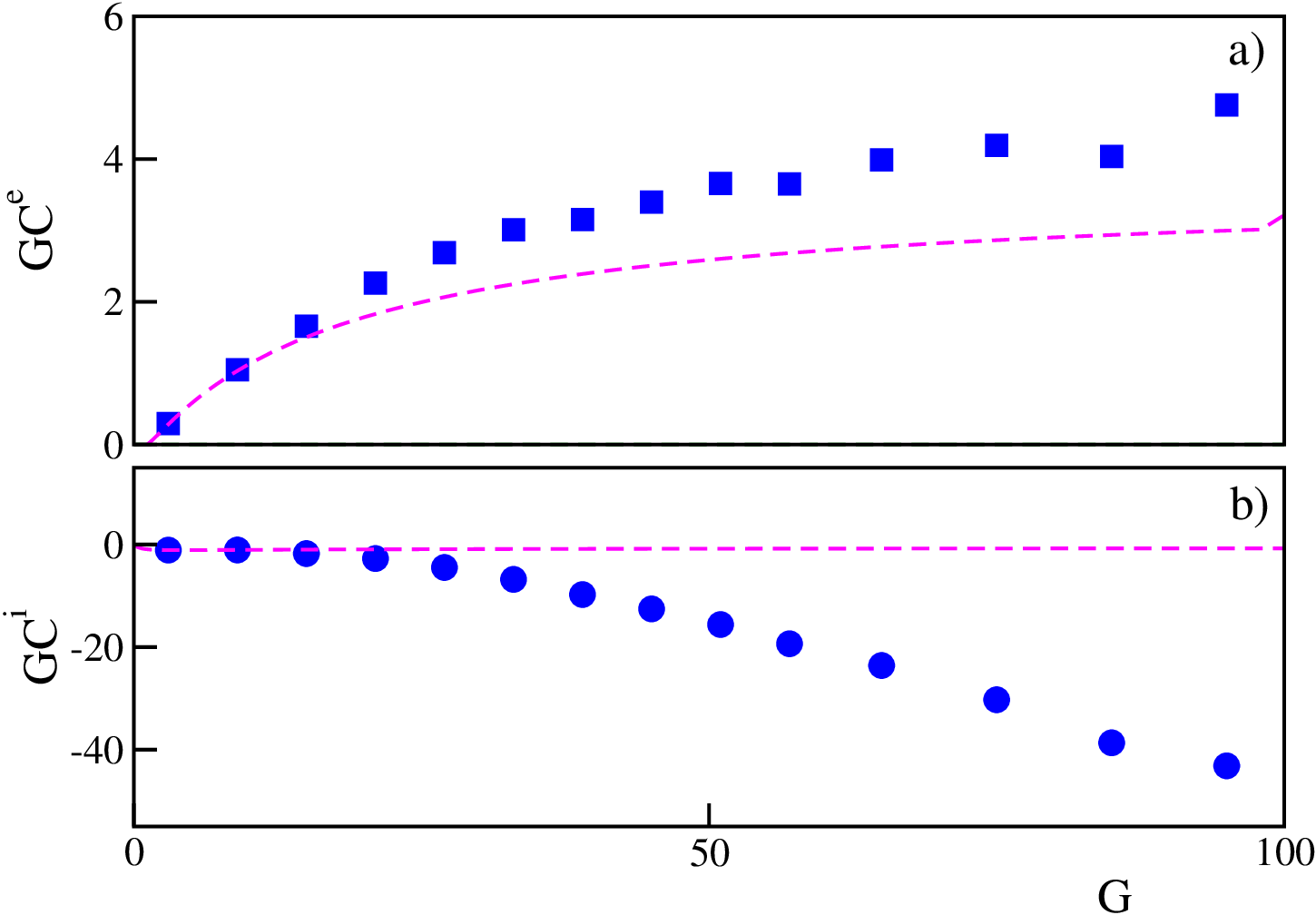}
\end{centering}
\caption{\label{fig:behavior} 
Aggregate synaptic currents $C^{e/i}$ multiplied by the coupling versus 
the synaptic coupling itself $G$ for excitatory (a) and inhibitory (b) neurons for $N=16000$.
The magenta dashed curves correspond to the exact calculations for the asynchronous regime.
 The parameters are as in Fig. \ref{fig:fields}. All the data refer to the annealed case.  
}
\end{figure}

The self-consistent results for the fields $E^{e/i}$, $I$ in the AR are presented in Figs.~\ref{fig:fields}(a), 
(magenta dashed lines), where we see that they reproduce the numerical observations
below the threshold $G_\theta$, in agreement with the expectations.
In the large $G$ limit, the AR still exists and the associated fields stay finite,
indicating that this regime is balanced. 
However, above $G_\theta$, the AR becomes unstable. The CC revealed by numerical simulations 
is also balanced, as visible in Fig.~\ref{fig:fields} (a), but
it is conceptually very different, not only because of the strong fluctuations of the fields.
This can be appreciated in Fig.~\ref{fig:behavior}(a-b), where the scaled 
aggregate currents $G C^{e/i}$ are plotted versus the coupling $G$ itself:
they remain bounded in the AR (magenta dashed lines), while they diverge in the presence of CC (solid symbols).

In the CC regime, the divergence of $GC^{e/i}$ visible in Fig.~\ref{fig:behavior}(a-b), seemingly contradicts the
finiteness of the fields themselves seen in Fig.~\ref{fig:fields} (a). The consistency can be restored by focusing on the role of the PRC. 
Typically, the efficacy of the incoming spikes, quantified by the PRC, is of order 1 irrespective of the coupling strength.
As shown in Fig.~\ref{fig:3}(a), where we 
plot the time and ensemble average $\langle Z^i_c \rangle$ of the PRC, conditioned to the times of 
the spike arrivals.
There, we see that $\langle Z^i_c \rangle$ decreases as $1/G$ for large $G$ (see black circles)
thereby compensating the effect of the increasing coupling strength (in the presence of finite currents).
This is an indirect, though clear, indication of an underlying synchronization, since $Z(\phi)$ is very small only in the vicinity of the threshold and the reset.

\begin{figure}
\begin{centering}
\includegraphics[width=0.5\textwidth,clip=true]{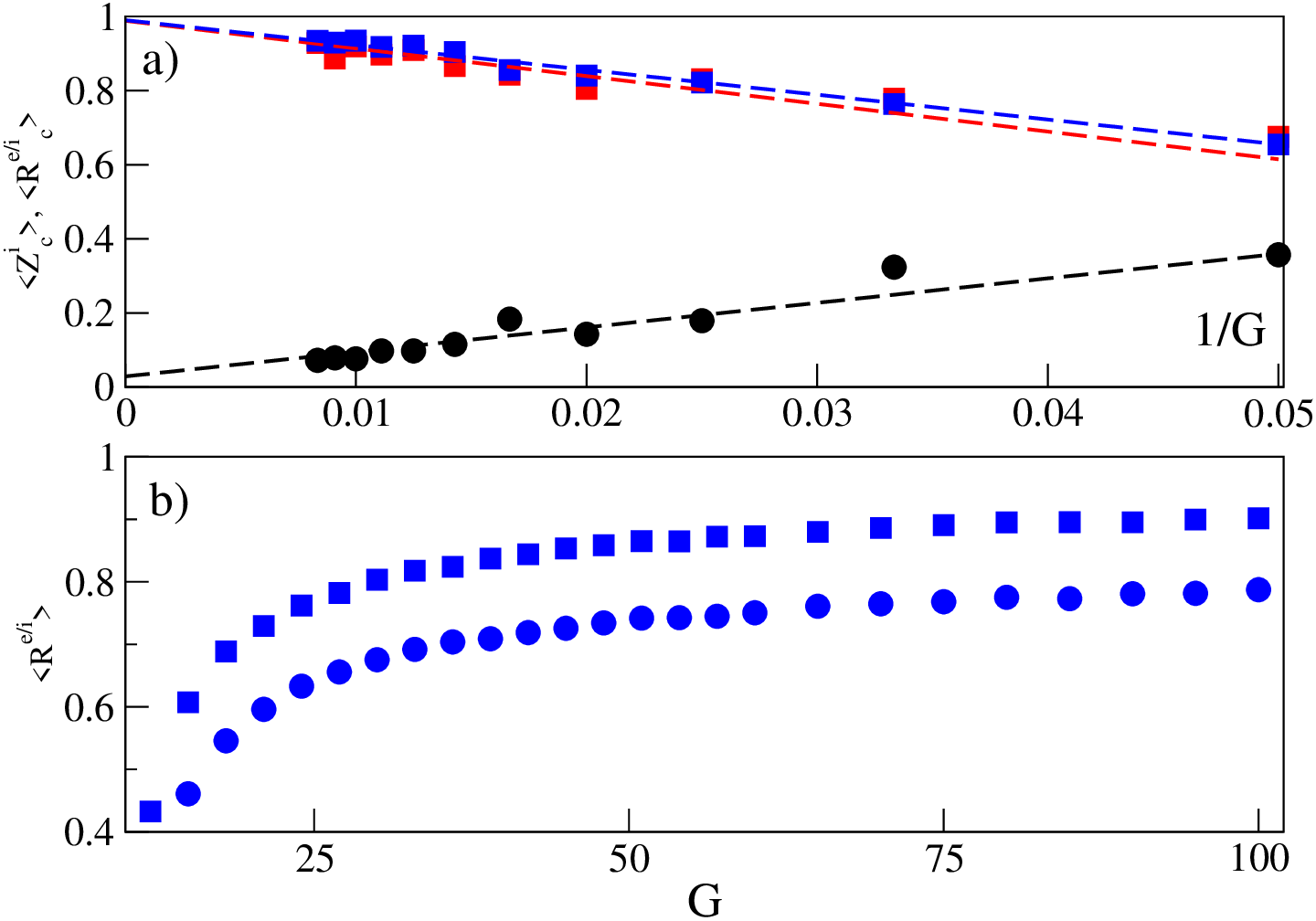}
\end{centering}
\caption{\label{fig:3} 
(a)  PRC and order parameters at EPSP arrival : 
$\langle Z^i_c \rangle$ (black circles), $\langle R^e_c \rangle$ (red squares) and
$\langle R^i_c \rangle$ (blue stars) versus $1/G$ for Z2 PRC.
We measure the average PRC of a typical inhibitory neuron whenever
a EPSP reaches it and at the same time we measure $R^e$ and $R^i$.
The dashed lines denote fiting to the data: namely, the black line
corresponds to $0.02888+6.6138/G$ for a fit to $\langle Z^i_c \rangle$; the red one to $0.9888-7.4705/G$
for a fit to $\langle R^e_c \rangle$
and the blue one to $0.9899 - 6.703/G$ for a fit to $\langle R^i_c \rangle$.
The data refer to  $N=16000$ and to an average over 250000 events. 
(b)  Average value of the Kuramoto order parameter $\langle R^{e/i} \rangle$ for the excitatory (circles) and inhibitory (squares)
population versus $G$ for $N=16000$.
Other parameters as in Fig. \ref{fig:fields}. All the data refer to the annealed case.}
\end{figure}
 
In Fig.~\ref{fig:3}(a) we show also the time averages $\langle R^{e/i}_c \rangle$ 
of the order parameters, conditioned to spike emissions, for the two
neural populations. In both cases, the parameter approaches one with a rate $1/G$, suggesting a convergence 
towards perfect synchronization when $G\to +\infty$.
However, this is not the whole story. In fact, the two unconditioned time averages 
$\langle R^{e/i} \rangle$  plotted in 
Fig.~\ref{fig:3}(b) remain strictly smaller than 1. 

Therefore, the overall scenario is not only qualitatively similar to the one observed for the quenched setup, but even quantitatively
close.

\subsection{Network Simulations}

The evolution equations Eq. (1) and (7) in the letter are integrated
by employing an event driven technique. Since they can be both splitted
in an exact evolution between two consecutive speike emissions and
and in a nonlinear update of the phases and of the synaptic efficacies
at each spike emissione, based on the actual value of the integrated
variables as well as for the phases on the values of the
the aggregate fields $C^{e/i}$ and of the PRC $Z(\phi_n^{e/i})$.
For the evolution of the synaptic efficacies we fix
the following parameters $u=0.5$ and $\tau_d = 1/0.35 = 2.8571\dots$.

The excitatory and inhibitory fields employed
in Fig. 1 (b-c) and in Fig. S1 (b-c) have been obtained by filtering the
spike trains with an exponential kernel. In practice, we
associate to each spike a an exponential pulse $p_\alpha(t) = \alpha {\rm e}^{- \alpha t}$ for $t>0$,
and the corresponding excitatory field $E^i$ can be obtained by integrating the following ODE:
\begin{equation}
{\dot E^i}(t) = -\alpha E^i(t) + \frac{\alpha}{N} \sum_{k, m|t^e(k,m)<t}  \delta(t-t^e(k,m))
\end{equation}
where $t^{e}(k,m)$ denotes the delivery time of the $m$-th spike by the $k$-th excitory neuron.
Analogously we have obtained the filtered inhibitory fields. We have always set $\alpha=10$.

\end{document}